\documentclass[12pt,fleqn]{article}

\usepackage{graphicx}
\usepackage{setspace}
\usepackage{amsmath}
\usepackage{amsthm}
\usepackage{amsfonts}
\usepackage{multirow}
\usepackage{booktabs}
\usepackage{amssymb}
\usepackage{color}
\usepackage{url}
\usepackage[comma]{natbib}
\usepackage{authblk}

\RequirePackage[comma]{natbib}

\def\BF{{\rm BF}}

\newcommand{\ev}{\mbox{\boldmath$e$}}

\newcommand{\gv}{\mbox{\boldmath$g$}}

\newcommand{\yv}{\mbox{\boldmath$y$}}
\newcommand{\Yv}{\mbox{\boldmath$Y$}}
\newcommand{\Yva}{\mbox{\boldmath{$\mathcal{Y}$}}}
\newcommand{\Zv}{\mbox{\boldmath{$\mathcal{Z}$}}}

\newcommand{\lv}{{\bf 1}}
\newcommand{\FDR}{\mbox{FDR}}
\newcommand{\FDP}{\mbox{FDP}}
\newcommand{\BFDR}{\overline{\mbox{FDR}}}

\newtheorem{prop}{PROPOSITION}
\newtheorem{lemma}{LEMMA}

\newtheorem{corollary}{COROLLARY}
\numberwithin{equation}{section}

\begin{document}
\title{\large \bf Robust Bayesian FDR Control using Bayes Factors, with Applications to Multi-tissue eQTL Discovery}
\author
{Xiaoquan Wen \\
Department of Biostatistics,  University of Michigan, Ann Arbor, USA}
\date{}
\maketitle
\begin{abstract}
Motivated by the genomic application of expression quantitative trait loci (eQTL) mapping, we propose a new procedure to perform simultaneous testing of multiple hypotheses using Bayes factors as input test statistics. One of the most significant features of this method is its robustness in controlling the targeted false discovery rate (FDR) even under misspecifications of parametric alternative models. Moreover, the proposed procedure is highly computationally efficient, which is ideal for treating both complex system and big data in genomic applications.
We discuss the theoretical properties of the new procedure and demonstrate its power and computational efficiency in applications of single-tissue and multi-tissue eQTL mapping.  
\end{abstract}

\section{Introduction}

With the emergence of high-throughput experiment technologies,  false discovery rate (FDR) control \citep{Benjamini1995} has become a standard statistical practice in many genomic applications.
Despite the importance of the FDR control procedures, the overall power in a multiple testing problem is largely determined by the test statistics employed. In many application areas, the practitioners are facing increasingly complex systems in which many factors must be accounted for, and their relationships are typically nontrivial. Consequently, it is extremely difficult to describe these systems mathematically without parametric modeling, and it is even more difficult to determine powerful testing statistics. 
Bayesian models have exhibited some great advantages in such complex settings and have become increasingly popular. Within the Bayesian hierarchical modeling framework, Bayesian FDR control is conceptually straightforward and easy to implement in a sound decision-theoretic framework, as illustrated by \cite{Newton2004, Mueller2006, Whittemore2007, Ji2008}. Nevertheless, its practical usage is often hindered primarily by two factors: first, its lack of robustness due to the heavy reliance on parametric assumptions in alternative model specifications \citep{Opgen2007} and second, its expensive computational cost.

This paper addresses both issues and proposes a novel Bayesian FDR control procedure motivated by the genomic application of expression quantitative trait loci (eQTL) mapping, where tens of thousands association tests are simultaneously performed. 
In the context of eQTL mapping, parametric models are particularly preferred because they provide necessary flexibility to i) control unknown confounding/batch effects; ii) incorporate prior knowledge of relevant genomic annotations; and iii) handle complex data structure, e.g., in multi-tissue eQTL mapping.
Those parametric models for eQTL mapping often contain hierarchical structures with natural Bayesian interpretations \citep{Flutre2013}. 
As a unique feature, our proposed procedure directly works with Bayes factors \citep{Kass1995}, which are often considered to be the natural test statistics within the Bayesian framework due to the likelihood principle. 
Most significantly,  our proposed procedure controls the target FDR level even when the alternative models are misspecified, which is a property that is currently missing from most available Bayesian and parametric model-based multiple testing approaches.   

In theory, it is always possible to convert a Bayes factor to a corresponding $p$-value (by treating the Bayes factor as a regular test statistic) and to apply the $p$-value based FDR control procedures \citep{Benjamini1995,Storey2002,Storey2003,Storey2003a, Storey2004, Storey2007}, a strategy known as the Bayes/Non-Bayes compromise \citep{Good1992}. 
However, in practice, the conversion often relies on permutation procedures and their computational cost can be extremely high when treating complex models and/or large data sets, therefore not suitable to apply to the genomic data at the genome-wide scale.
In fact, the computational efficiency is a general challenge for any multiple testing procedure that relies on permutation $p$-values. 
In comparison, for a wide range of parametric models, it has been demonstrated that Bayes factors can be computed or accurately approximated in a highly efficient manner \citep{Kass1995, Raftery1996, Diciccio1997, Johnson2005, Servin2007, Liang2008, Johnson2008, Saville2009,  Flutre2013, Wen2014a, Wen2014b}. Based on these results, the computational resources required by our proposed procedure for controlling FDR are trivial, which makes it ideal for treating both complex models and big data in genomic applications.

The remainder of the paper is organized as follows: we first provide the necessary background regarding FDR controls within the Bayesian framework in the Section 2, and then proceed to describe our novel Bayesian FDR control method in the Section 3. In the Section 4 and 5, we illustrate the proposed approach in applications of single-tissue and multi-tissue eQTL mapping, respectively. 
Finally, we summarize our findings and discuss future directions in the Section 6.

\section{Notation and Setup}

Throughout this paper, we consider  multiple testing problems under a general hierarchical mixture model setting, which is generally assumed in both Bayesian and frequentist frameworks \citep{Efron2001,Storey2003,Newton2004,Genovese2004, Mueller2006}. 
Consider a set of $m$ hypotheses for simultaneous testing. Let $\Yv_i$ denote the data collected for the $i$-th test, and let
the complete collection of the observed data from all $m$ tests be denoted by $\Yva := \{ \Yv_1,\dots, \Yv_m\}$.  Suppose that each $\Yv_i$ is generated from either a null model or some alternative model, namely,
\begin{equation}
   \Yv_i \sim p_i^{Z_i},
\end{equation}  
where $Z_i$ is a latent indicator that denotes the true generating distribution. 
Note that we do not require all the hypotheses to share the same null or alternative models. 
The vector of the latent indicators, $\Zv := (Z_1, \dots, Z_m)$, is of primary interest for inference, and the commonality of all the tests lies on the assumption
\begin{equation}
  Z_i \mid \pi_0 \,  \stackrel{i.i.d}{\sim} \, {\rm Bernoulli}\,\left(1-\pi_0\right),~i=1,...,m.
\end{equation}
In particular, the parameter $\pi_0$ represents the proportion of the data generated from the respective null hypothesis and is typically unkonwn.
Finally, we assume that the tests are mutually independent conditional on $\pi_0$, i.e.,
\begin{equation}
  \Pr(\Zv \mid \Yva, \pi_0) = \prod_{i=1}^m \Pr(Z_i \mid \Yva, \pi_0) = \prod_{i=1}^m \Pr(Z_i \mid \Yv_i, \pi_0).
\end{equation} 

Under the above hierarchical mixture model, the simultaneous hypothesis testing can be framed as a {\em joint} decision problem with respect to $\Zv$.
Let $\delta_i$ denote a decision (0 or 1) on $Z_i$ based on all the observed data, and define $D:= \sum_{i=1}^m \delta_i$. 
Following the formulation of \cite{Mueller2006}, we define the False Discovery Proportion (FDP) as the proportion of false discoveries among total discoveries, i.e., $
 \FDP := \frac{\sum_{i=1}^m   \delta_i (1-Z_i)}{D \vee 1}$.
The Bayesian False Discovery Rate is naturally defined as the expectation of FDP conditional on $\Yva$, i.e.,
\begin{equation}
  \label{bfdr.def}
 \BFDR := {\rm E}(\FDP \mid \Yva) =  \frac{\sum_{i=1}^m   \delta_i \left(1- v_i \right)}{D \vee 1},
\end{equation}
where the conditional expectation is taken with respect to $\Zv$, and $v_i := \Pr(Z_i=1 \mid \Yva)$.
Moreover, the frequentist control of the False Discovery Rate focuses on the quantity
\begin{equation}
  \FDR := {\rm E}(\BFDR) = {\rm E}\left[{\rm E}(\FDP \mid \Yva)\right], 
\end{equation}
where the additional expectation is taken with respect to $\Yva$ over (hypothetically) repeated experiments.
It is important to note that controlling the Bayesian FDR is a {\em sufficient but not necessary} condition to control the corresponding frequentist FDR; thus the Bayesian FDR control is more stringent in theory.
For measuring type II errors in a multiple testing setting, the False Non-discovery Proportion (FNP), the Bayesian False Non-discovery Rate (FNR) and the frequentist FNR are defined in a similar fashion.

Bayesian FDR control procedure is carried out by evaluating the posterior probability $v_i$ for each test, i.e.,
\begin{equation}
  \label{bayes.post}
  v_i = \int \Pr(Z_i = 1 \mid \Yva, \pi_0)\, p (\pi_0 \mid \Yva)\,d\, \pi_0.
\end{equation} 
In particular, the data from all tests jointly affect $p(\pi_0 \mid \Yva)$ and thereby impact the posterior distribution of each individual $Z_i$. 
However, conditional on $\pi_0$, $\Pr(Z_i = 1 \mid \Yva, \pi_0)$ can be computed independently, i.e.,
\begin{equation}
  \label{lfdr}  
   \Pr(Z_i = 1 \mid \Yva, \pi_0) =  \Pr(Z_i = 1 \mid \Yv_i, \pi_0) = \frac{(1-\pi_0)\, \BF_i}{\pi_0 + (1-\pi_0)\, \BF_i},
\end{equation}
where $\BF_i$ is the null-based Bayes factor \citep{Liang2008} defined by 
\begin{equation}
  \label{bf.def}
\BF_i := \frac{ p(\Yv_i \mid Z_i = 1)}{p(\Yv_i \mid Z_i = 0)} = \frac{p^1_i(\Yv_i)}{p^0_i(\Yv_i)}.
\end{equation}
Particularly, in computing $\BF_i$, all the potential nuisance parameters in the parametric models (null or alternative) are integrated out with respect to their prior distributions \citep{Kass1995}.  
As demonstrated by \cite{Newton2004} and \cite{Mueller2006}, the Bayesian FDR control is based on the following natural decision rule,
\begin{equation}
  \label{decision.rule}
  \delta_i^*(t) = I \left(v_i > t \right),
\end{equation}
i.e., the null hypothesis is rejected if the posterior probability that the observed data is generated from the alternative is high. 
For a pre-defined FDR level $\alpha$, the threshold $t_\alpha$ in (\ref{decision.rule}) is determined by  
\begin{equation}
 t_{\alpha} =   \arg\min_t  \left( \frac{\sum_{i=1}^m  \delta_i^*(t) (1-v_i )}{D(t) \vee 1} \le \alpha \right)
\end{equation}
Furthermore, \cite{Mueller2004} demonstrated that the decision rule (\ref{decision.rule}) is optimal in the sense that it minimizes the corresponding FNR while controlling for the FDR.

\section{Robust Control of Bayesian FDR}

Although the Bayesian FDR control is conceptually straightforward, its practical performance is susceptible to {\em alternative} model misspecifications. In comparison, the $p$-value based frequentist FDR control procedures demand only adequate behavior of $p$-values under the null models and generally ensure targeted FDR control levels, regardless of the distribution of $p$-values under the assumed alternatives.  
This robustness property against type I error is often desirable in the hypothesis testing context.

In this section, we provide a detailed discussion of this topic and provide intuitions and theoretical justifications for a new Bayesian FDR control procedure. It should be noted that our notion of `` model misspecification" includes the cases of prior misspecification in Bayesian settings.

\subsection{General Approach}

The specifications of parametric (null and alternative) models impact Bayesian FDR control in the following two aspects. 

First, the computation of the Bayes factors relies on the model assumptions of $p_i^0$ and $p_i^1$. In hypothesis testing, the ``correctness" of Bayes factors is formally characterized by a concept known as {\em consistency} in model comparisons, i.e., with sufficiently large sample size, Bayes factor $\to 0$ if the data is truly generated from the null model and $\to \infty$ if the data is generated from the alternative model.  
Intuitively, to prevent inflation of the pre-defined FDR level, which concerns the behavior of the testing procedure under the null, the consistency of the Bayes factors is only required under the null models. Such a condition is easy to satisfy (and also easy to check) as long as the null model is correctly specified.
In contrast, when the alternative model is misspecified (which is almost certain in practice), the Bayes factor might not be consistent under the alternative. Nonetheless, as we will demonstrate later, it does not inflate the FDR but only affects the power/FNR {\em given the true $\pi_0$}.
We also note that the assumption regarding the consistency of the Bayes factors under null models is analogous to the uniformity assumption of the null $p$-values, i.e., both are valid only if the null models are correctly specified. 
In summary, we expect that the null models are correctly specified and allow imperfect specification of the alternative models . 
Nevertheless to ensure decent power in testing, it is preferred that the Bayes factor is far greater than 1 under the alternative scenario and especially in an asymptotic scenario. 
Importantly, the consistency property of the Bayes factors under the null models implies that the Bayes factors from the alternative models are stochastically dominant over the Bayes factors from the null models.

Second, the inference of $p(\pi_0 \mid \Yva)$  from the mixture model is sensitive to the alternative model specifications and therefore affects the Bayesian FDR control: if $\pi_0$ is systematically under-estimated, the pre-defined Bayesian FDR level is inflated; conversely, over-estimating $\pi_0$ results in a conservative procedure that loses power but maintains the desired FDR level. 
To avoid under-estimation of $\pi_0$ in all circumstances, we attempt to find a robust upper bound estimator for $\pi_0$, denoted by $\hat \pi_0 (\Yva)$, such that under the sampling distribution of $\Yva$ and for all possible values of $\pi_0$,
\begin{equation}
  \label{conserv.prop}
   \Pr\left(\hat \pi_0 (\Yva) \ge \pi_0 \mid \pi_0 \right) \xrightarrow{} 1, ~ \text{as the number of the true null tests is sufficiently large}. 
\end{equation}
The conditional probability expression emphasizes that the upper bound property of $\hat \pi_0(\Yva)$ holds true for arbitrary $\pi_0$ values. 
Given (\ref{conserv.prop}), for arbitrary prior distributions on $\pi_0$, it follows that
\begin{equation}
   \Pr( \pi_0 \le \hat \pi_0(\Yva) \mid \Yva) \xrightarrow{} 1,
\end{equation}
i.e., $\hat \pi_0$ forms a probability upper bound of the posterior distribution of $\pi_0$, whose corresponding prior distribution can be arbitrary.

The key to our approach is to ensure condition (\ref{conserv.prop}) even if the alternative models are misspecified. (It should be clear that such an upper bound does exist: e.g., the trivial bound $\hat \pi_0(\Yva) \equiv 1$ satisfies this requirement; however, it is too conservative to be useful.) 
In this paper, we rely on results from asymptotic theories to find the desired $\hat \pi_0$ estimates while utilizing virtually no assumptions from the alternative models.
As a consequence, the resulting $\hat \pi_0$ may not be optimal in terms of ``precision" (with respect to the truth), but it certainly has great appeal in guarding the pre-defined FDR level against alternative model misspecifications. 

Based on the above discussion, we now define a new quantity
\begin{equation}
  \hat v_i := \Pr(Z_i = 1 \mid \Yv_i, \hat \pi_0) = \frac{(1-\hat \pi_0) \BF_i }{\hat \pi_0 + (1-\hat \pi_0) \BF_i},
\end{equation}
which is consistently more conservative than the underlying true $v_i$.  We propose a conservative Bayesian FDR control procedure that essentially replaces $v_i$ with  $\hat v_i$ in (\ref{decision.rule}). 
Specifically, we denote the new decision rule $\delta_i^\dagger(t) := I \left( \hat v_i > t \right)$ and $D^\dagger(t) := \sum_{i=1}^m \delta_i^\dagger (t)$. 

With a stronger assumption that assumes sufficiently large sample size for each test, we are able to show the following theoretical result for the proposed approach:
\begin{prop}
Assuming that the following conditions are satisfied:
 \begin{enumerate}
    \item the Bayes factors are consistent under the true null models   
    \item there exists an upper bound  point estimator $\hat \pi_0(\Yva)$ such that 
    $$ \Pr(\pi_0 \le \hat \pi_0 (\Yva) \mid \pi_0) \to 1, \forall \, \pi_0 \in (0,1),$$ as the number of the true null tests is sufficiently large. 
 \end{enumerate}
The decision rule $\delta_i^\dagger(t) := I \left( \hat v_i > t \right)$ controls the frequentist FDR at level $\alpha$ with the rejection threshold
 $$  t^\dagger_{\alpha} =   \arg\min_t  \left( \frac{\sum_{i=1}^m   \delta_i^\dagger(t) (1-\hat v_i )}{D^\dagger(t) \vee 1} \le \alpha \right),$$
as the sample size per test and the number of the true null tests are sufficiently large.
Furthermore, the Bayesian FDR can be consistently controlled in the sense that
\begin{equation*}
  \Pr \left( \overline{\rm FDR} \le \alpha \right) \xrightarrow{} 1.
\end{equation*} 
\end{prop}

The proof is based on the above discussion and the full details are given in the appendix A.
Proposition 1 specifies two asymptotic conditions: both the number of true null tests and the sample size per test are required to be large. 
The large number of the true null tests ensures the probabilistic nature of the upper bound property of $\hat \pi_0$. This condition is also implicitly required for $p$-value based FDR control procedures to guarantee the uniform distribution of $p$-values from the true null tests. (This point is demonstrated by the connection between the QBF procedure and the Storey's $q$-value procedure in the section 3.2.2). 
It is important to note that this condition does not impose any restriction on the proportion of the true null tests, i.e., $\pi_0$.
The condition for sample size per test is used for proving the stringent Bayesian FDR control criteria.
Although we believe that the result of the Proposition 1 holds true even without this particular large sample requirement (especially for controlling frequentist FDR), the proof becomes extremely difficult and our attempt is unsuccessful. 
Nevertheless in practice, we observe that the stated decision rule works well for very modest sample sizes: illustrations are given in the simulation study and the real data application where the sample sizes are both quite limited. 

\subsection{Upper Bound Estimation of $\pi_0$ using Bayes Factors}

In this section, we describe two methods that can robustly estimate $\hat \pi_0$ that satisfy assumption 2 in the Proposition 1 using only Bayes factors. 
The two methods share a simple intuition based on the stochastic orderings: for Bayes factors with reasonable powers, the smaller values are more likely generated from the null models. Theoretically, they both rely on the weak law of large numbers (WLLN) to ensure the upper bound property stated in (\ref{conserv.prop}). Note that both approaches require the number of the simultaneous tests is sufficiently large, but impose no restrictions on sample sizes in each test.   
R code code implementing the described computational procedures is made freely available at \url{https://github.com/xqwen/bfdr}.

\subsubsection{Sample Mean based Estimator}
We first introduce an estimation procedure for $\hat \pi_0$ using the sample mean of the observed Bayes factors.  This procedure, named the EBF-procedure, is based on the following Lemma:
\begin{lemma}
A well-defined Bayes factor satisfies 
  \begin{equation*}
   {\rm E}( {\rm BF} \mid H_0) = 1.
  \end{equation*}
\end{lemma}
The proof of the Lemma is trivial from the definition of the Bayes factor. 

Intuitively, under some suitable conditions and by the large sample theory, the Lemma 1 implies that the sample means of the Bayes factors from the true null models converge to 1. On the other hand, the Bayes factors from the true alternative models with reasonable powers should be, on average, greater than 1 (i.e., favoring the alternative over the null models). Therefore, the sample mean of the observed Bayes factors carries information regarding the mixture percentage.       
We give a detailed description of the EBF procedure below:
\begin{enumerate}
  \item Sort all observed Bayes factors in ascending order, 
    \begin{equation*}
    \BF_{(1)}, \BF_{(2)}, \dots, \BF_{(m)}
    \end{equation*}
  \item Find 
 \begin{equation}
    d_0 = \arg \max_{d} \left( \frac{1}{d} \sum_{i=1}^d {\rm BF}_{(i)} < 1 \right)
 \end{equation}
 \item Estimate $\hat \pi_0$ by
 \begin{equation}
     \hat \pi_0 = \frac{d_0}{m}
 \end{equation}
\end{enumerate}  
Essentially, this procedure searches for the largest proportion of tests with mean Bayes factors less than $1$. 
We demonstrate in Lemma 2 and Proposition 2 that the EBF procedure produces the upper bound estimate of $\pi_0$ required by the Proposition 1.

\begin{lemma}
In the case $\pi_0 = 1$, the EBF procedure estimates $\hat \pi_0 \xrightarrow{P} 1$,  provided that the number of tests is sufficiently large. 
\end{lemma}
The proof is simply based on the law of large numbers, and the full details are given in the appendix \ref{lem2.proof}.  
\begin{prop}
For a mixture of null and alternative models, the estimate by the EBF procedure satisfies
  $$ \Pr( \hat \pi_0 \ge \pi_0 \mid \pi_0 ) \to 1, $$
provided that the number of the null tests is sufficiently large.
\end{prop}
The Proposition 2 is trivially followed from Lemma 2, and the detailed argument is provided in the appendix \ref{prop1.proof}.

The EBF procedure also allows the study of some of the general properties of Bayesian FDR control. Here, we show one interesting example: if a single Bayes factor exceeds a certain threshold, the corresponding null hypothesis can be rejected with respect to a targeted FDR level without the explicit evaluation of $\hat \pi_0$. We call this threshold the {\em automatic rejection threshold} of Bayes factors in multiple testing. We have the following corollary concerning one such automatic rejection threshold:
\begin{corollary}
Under the settings stated in Proposition 1 and at the FDR level $\alpha$, if a Bayes factor exceeds $m/\alpha$, the EBF procedure automatically rejects the corresponding null hypothesis.
\end{corollary} 
The proof is trivially based on the estimate of $\hat \pi_0$ by the EBF procedure and equation (\ref{lfdr}); we give the details in the appendix D. Most interestingly, the automatic rejection threshold, $m/\alpha$, clearly resembles Bonferroni corrections for $p$-values.

\subsubsection{Sample Quantile based Estimator }\label{qbf.sec}

Alternatively, we can estimate $\hat \pi_0$ using the sample quantile information based  on the distribution of Bayes factors under the null models. 
Let $F_i^0(x)$ and $F_i^1(x)$ denote the cumulative distribution functions of $\BF_i$ under the null and the true (unknown) alternative models, respectively. Suppose that
\begin{equation}
  F_i^0(q_{i,\gamma}) = \gamma,
\end{equation}
i.e., $q_{i,\gamma}$ is the $\gamma$-quantile of $\BF_i$ under the null model. 
Consider the sample mean of the independent indicators $I (\BF_i \le  q_{i,\gamma})$, for $i = 1,...,m$. By Chebyshev's weak law of large numbers, it follows that
\begin{equation}  
 \label{qlln}
 \frac{1}{m} \sum_{i = 1}^m I(\BF_i \le q_{i,\gamma})  \xrightarrow{P}   \pi_0 \gamma + \frac{1-\pi_0}{m} \sum_{i=1}^m  F_i^1(q_{i,\gamma}).
\end{equation}
 Note that if $\gamma$ is small and/or $F_i^0(x)$ and $F_i^1(x)$ are well separated, the contribution from the $F_i^1(q_{i,\gamma})$ terms should also be small. 
This is because Bayes factors computed using data from the alternative models are stochastically dominant over the Bayes factors evaluated using data generated from the null models, a property implied by the consistency of the Bayes factors under the null models.
Based on (\ref{qlln}), we propose an estimating algorithm, named the QBF procedure, to estimate $\hat \pi_0$. We summarize the details of the QBF procedure and its theoretical properties in the following proposition:
\begin{prop}
For a given $\gamma$, the QBF-procedure estimates
\begin{equation}\label{hpi0.q}
    \hat \pi_0 = \frac{\sum_{i = 1}^m I(\BF_i \le  q_{i,\gamma })}{m \gamma}.
\end{equation}
As the number of the null tests is sufficiently large, it follows that
\begin{equation*}
   \Pr(\hat \pi_0 \ge \pi_0 \mid \pi_0) \to 1.
\end{equation*}
\end{prop}
The proof follows trivially from (\ref{qlln}) by ignoring the contribution from the term $\frac{1-\pi_0}{m} \sum_{i=1}^m  F_i^1(q_{i,\gamma}) $ on the right-hand side. 

It should be noted that the QBF-procedure is connected with what was proposed by \cite{Storey2004}:  treating $\BF_i$ as a test statistic and denoting its $p$-value by $p_i$, under the null model, it follows that, 
\begin{equation}
   \label{bf.p.con}
   \BF_i \le q_{i,\gamma}  ~ \Leftrightarrow ~ p_i > 1 - \gamma.
\end{equation} 
This yields an equivalent representation of (\ref{hpi0.q}):
\begin{equation}
    \hat \pi_0 = \frac{\sum_{i = 1}^m I(p_i >  1 - \gamma )}{m \gamma},        
\end{equation}
which is the exact formula proposed in \cite{Storey2004}.

Based on (\ref{qlln}), it can be concluded that the QBF procedure generates an upper bound  estimate of $\pi_0$ using any $\gamma \in (0,1)$ asymptotically. However, for Bayes factors with reasonable power, taking smaller $\gamma$ values yields relatively tighter upper bounds on $\pi_0$ (because of the relatively small contribution from $F_i^1(q_{i,\gamma})$). 
However if $\gamma$ is too small, the estimator $\hat \pi_0$ shown in (\ref{hpi0.q}) has undesired large variance for finite $m$. 
To strike a balance, we simply set $\gamma=0.5$ for demonstrations in this paper. (\cite{Storey2004, Storey2003a} considered the choice of $\gamma$ as a problem of bias-variance tradeoff and offered an elegant statistical solution to find an ``optimal" $\hat \pi_0$ by using multiple $\gamma$ values. It should be noted that their solution also works in our case.) 

When the quantiles of the Bayes factors under the null models are not directly available, we can always employ a permutation procedure to estimate them from the samples at hand. According to (\ref{bf.p.con}), the permutation procedure is no different statistically than computing permutation $p$-values. 
However, because the quantiles preferred by the QBF procedures are typically not in the tail area, they can be estimated stably with fewer permutations (than is required for accurately assessing permutation $p$-values).
Therefore, even in this case, the QBF procedure is still computationally efficient. We will fully illustrate this point in our numerical studies.

\subsubsection{Comparison of EBF and QBF Procedures}

One of the practical differences between the QBF and EBF procedures lies in the relative conservativeness of the resulting $\hat \pi_0$ estimates. We observe that the QBF procedure with $\gamma = 0.5$ typically yields a tighter upper bound than the EBF procedure. This might be explained by the choice of parameter $\gamma$ in the QBF procedure:  smaller $\gamma$ values yields tighter upper bound estimates. Although the EBF procedure does not explicitly use the quantile information, computing a sample mean from the null distribution of the Bayes factors requires examining $\BF_i \le q_{i,\gamma}$ for $\gamma \to 1$, i.e., nearly the full distribution is used, and this is seemingly analogous to applying the QBF procedure with a large $\gamma$ value.

Once the Bayes factors are computed, the computational cost for the EBF procedure is trivial in all cases. However, for the QBF procedure, in many situations in which the quantiles of the Bayes factors under the null models are not directly computable, there is an additional cost of permutations for quantile estimations.

\section{Simulation Study: Single-tissue eQTL Discovery}

In this section, we demonstrate the proposed robust Bayesian FDR control method in the context of eQTL mapping, which aims to identify genetic variants that are associated with gene expression levels.
Specifically, we consider a problem of genome-wide eQTL mapping in a single tissue for which we can compare the proposed Bayesian approach with  the gold-standard frequentist approach based on permutation $p$-values .

Particularly, we consider the expression levels of $L$ different genes across the genome are measured. 
For each gene $i$, we consider $k_i$ candidate SNPs, which are typically located in the {\it cis} region of the target gene, and model its expression level $\yv_i$ in the sample using the following multiple linear regression model, 
\begin{equation}
  \label{sim2.model}
  \yv_i = \mu_i \lv + \sum_{j=1}^{k_i} \beta_{i_j} \gv_{i_j} + \ev_i,~~\ev_i \sim {\rm N}( 0, \sigma_i^2 I),
\end{equation}
where $\gv_{i_j}$ denote the genotypes of the $j$-th  candidate genetic variant in gene $i$, and the $\gv_{i_j}$s are typically correlated. 
Based on the linear model, the problem of eQTL discovery can be framed as testing the following hypothesis for each gene $i$:
\begin{equation*}
   H_0: \beta_{i_j} = 0, \forall j ~ ~~{\rm vs.}~~~H_1: \mbox{ some  } \beta_{i_j} \neq 0,
\end{equation*}  
and we perform the simultaneous hypothesis testing for all $L$ genes.
In practice, this QTL discovery procedure for identifying genes harboring causal associated SNPs in their {\it cis} regions is referred to as {\em eGene discovery}.

\subsection{Simulation Scheme}
We take the real genotype data from a published study \citep{Barreiro2012} in which $n=85$ unrelated individual samples are genotyped genome-wide. We select 10,000 genes from the data set, where each gene contains 40 to 120 {\it cis}-SNPs. 
Within each gene, genotypes of different genetic variants are typically correlated at various levels due to linkage disequilibrium (LD). 
For each gene, we simulate its phenotype vector according to the linear model (\ref{sim2.model}). More specifically, we fix the intercept term to $\mu_i = 1$ and residual error variance to $\sigma^2 = 1$, and we again assume that it is known. Under the alternative scenario, for each gene, we randomly sample one to five associated SNPs , and simulate their genetic effect sizes by drawing from the distribution $\beta_{i_j} \sim {\rm N}(0, 0.6^2)$.  
For each gene, with probabilities $\pi_0$ and $1-\pi_0$ we simulate under the null and the alternative scenarios, respectively.

\subsubsection{Methods for Comparison}

The testing problem encountered in this simulation setting is challenging, primarily because there is no straightforward test statistic available for the desired gene-level test. 
To obtain a valid gene-level $p$-value for gene $i$, the standard frequentist approach for eQTL mapping performs $k_i$ single SNP association tests and obtain $p$-values based on the resulting $t$-statistics.
Subsequently, the minimum of the $k_i$  $p$-values is regarded as the test statistic for the gene $i$. (We shall term this test statistic as the min-$p$ statistic). 
Intuitively, the min-$p$ statistic is most powerful if there is only a single associated SNP within the candidates \citep{Cruz2010}.
Note that the min-$p$ statistic itself is no longer a valid $p$-value, as it does not follow a uniform distribution under the null.
Furthermore, because of the complex correlation structures among the genotypes, the null distribution of the min-$p$ statistic is generally unknown. 
In practice, this issue is resolved by permuting the individual labels of the phenotype vectors and assessing a permutation $p$-value of the min-$p$ statistic, $p_{minp}$, for each gene. 
These gene-level $p$-values are then plugged into either the Storey procedure \citep{Storey2002, Storey2003} or the Benjamini-Hochberg (B-H) \citep{Benjamini1995} procedure for multiple testing control.

For Bayesian hypothesis testing, we apply a simple Bayesian model proposed by \cite{Servin2007}. 
Briefly, this model explicitly assumes that, under the alternative scenario, exactly one genetic variant is associated with the expression level, which is similar to the implicit assumption made by the standard frequentist approach.
Furthermore , it assumes that every candidate genetic variant is equally likely to be associated {\it a priori}. 
Under these assumptions, for gene $i$, there are exactly $k_i$ possible alternative single SNP association  models.
For each alternative model with assumed causal SNP $i_j$, 
a Bayes factor, $\BF_{i_j}$ can be analytically computed  as a simple function of the $t$-statistic, $\hat \beta_{i_j}/{\rm se}(\hat \beta_{i_j})$, by fitting the corresponding simple linear regression models \citep{Cox1979, Servin2007, Wakefield2009}.
We then compute a gene-level Bayes factor by averaging over all single SNP association models, i.e.,
\begin{equation}
 \overline {\BF}_i = \frac{1}{k_i} \sum_j {\BF}_{i_j}.
\end{equation}      
Note that $\overline {\BF}_i$ is  a well-defined Bayes factor, and can be directly used in the EBF and the QBF procedure.  

We apply both the EBF and the QBF procedures to analyze the simulated eQTL data. Because the null distribution of the Bayes factor is not available for the QBF procedure,  we use permutations to estimate the median of the null Bayes factor distribution for each gene. 
In addition,  we also obtain a permutation $p$-value by treating $\overline {\BF_i}$ as a gene-level test statistic, which is denoted by $p_{bf}$, and use it for the B-H and the Storey procedures.

\subsection{Simulation Results}

We simulate data for $\pi_0 = 0.15,...,0.95$ and create 20 independent replicates for each $\pi_0$ value. We apply the Storey and B-H procedures using both $p_{minp}$ and $p_{bf}$, and use the gene-level Bayes factors for the EBF and QBF procedures. 

We first examine the $\hat \pi_0$ estimates from the EBF, QBF, and Storey procedures. The estimates from the QBF and the Storey procedures are based on permutations. We vary the number of permutations from 100 to 500 to 5000 for $\gamma = 0.5$.
We average the $\hat \pi_0$ estimates over replicates and summarize the results in Table \ref{pi0.est.sim2.tbl}. 
The $\hat \pi_0$ estimates obtained by the Storey procedure based on $p_{bf}$ are almost identical to the QBF estimates with the same permutation repeats. To avoid the redundancy, they are not shown.
All the estimates satisfy the upper-bound requirement. The EBF procedure consistently yields the most conservative estimate. The results from the QBF and Storey procedures based on $p_{minp}$ are mostly similar, although the QBF estimates are slightly less conservative.
One important observation is that for all three methods, the $\hat \pi_0$ estimates have very little variation according to the number of permutations, which is mostly expected for the reason discussed in section \ref{qbf.sec}. 

\begin{table}[ht]
\begin{center}
\begin{tabular} { c   c  c c c  c  c c c  c   c }
\hline
   ~ & ~& \multicolumn{9}{c}{$\hat \pi_0$} \\
   \cline{3-11} 
  $\pi_0$ & ~& \multicolumn{3}{c}{Storey ($p_{minp}$) } & ~ & \multicolumn{3}{c}{QBF} & ~ & ~ \\
  \cline{3-5} \cline{7-9}
  ~ & ~ & 100 & 500 & 5000 & ~ & 100 & 500 & 5000 & ~ & EBF \\
\hline
\hline
 0.95 & ~ & 0.978 & 0.970 & 0.974 & ~ & 0.977 & 0.969 & 0.973 & ~& 0.982 \\
 0.85 & ~ & 0.914 & 0.910 & 0.905 & ~ & 0.911 & 0.907 & 0.903 & ~& 0.943 \\
 0.75 & ~ & 0.849 & 0.842 & 0.846 & ~ & 0.842 & 0.845 & 0.838 & ~& 0.901 \\ 
 0.65 & ~ & 0.781 & 0.784 & 0.787 & ~ & 0.772 & 0.775 & 0.779 & ~& 0.857 \\
 0.55 & ~ & 0.726 & 0.723 & 0.723 & ~ & 0.718 & 0.714 & 0.713 & ~& 0.813 \\
 0.45 & ~ & 0.659 & 0.660 & 0.657 & ~ & 0.647 & 0.650 & 0.647 & ~& 0.766 \\
 0.35 & ~ & 0.602 & 0.600 & 0.596 & ~ & 0.590 & 0.587 & 0.584 & ~& 0.718 \\
 0.25 & ~ & 0.532 & 0.535 & 0.530 & ~ & 0.517 & 0.520 & 0.514 & ~& 0.668 \\
 0.15 & ~ & 0.471 & 0.468 & 0.466 & ~ & 0.455 & 0.452 & 0.451 & ~& 0.616 \\              

\hline
\end{tabular}
\caption{\label{pi0.est.sim2.tbl} $\hat\pi_0$ estimates by the Storey,  QBF, and EBF procedures in the simulation study. Each entry indicates the average estimates for different $\pi_0$ values  taken over 20 simulated data sets. For the Storey and the QBF procedures, the estimates are based on the permutation procedures; the results obtained for different permutation numbers are also shown. All estimates satisfy the upper bound requirement and they are not very sensitive to the number of permutations.}

\end{center}
\end{table}

We then proceed to examine the realized false discovery and false non-discovery rates obtained by the B-H,  Storey,  QBF, and EBF procedures. The results are summarized in Tables \ref{fdp.sim2.tbl} and \ref{fnp.sim2.tbl}. 
Because the estimates of $\hat \pi_0$ from the QBF procedure have only small variation for different numbers of permutations, its realized FDP and FNP results are virtually invariant with respect to the permutation numbers; therefore, we only report the results using $\hat \pi_0$ estimated by 100 permutations. 
Contrary to the Bayesian procedures, the number of permutations directly impacts the precision of the permutation $p$-values and consequently alters the outcomes of the $p$-value based FDR control procedures. Such impacts can be severe especially when the value of $\pi_0$ is close to 1. In our experiment, when $\pi_0 = 0.95$, with 500 permutations, neither the Storey nor the B-H procedure is able to reject a single null hypothesis. 
The issue seems to be resolved by performing more permutations; however, we have to pay a steep price in computations even in this simple setting.

Overall, all the methods maintain the desired level of FDR control, and their powers are quite comparable (when the accurate $p$-values are provided to the Storey and B-H procedure). More specifically, we note that
\begin{enumerate}
  \item Results from the B-H and  Storey procedures based on $p_{bf}$ provide benchmarks to compare the frequentist and the Bayesian FDR controls in this setting. The Bayesian FDR controls are expected to be more stringent, but the differences reflected by the realized FNPs are mostly small (especially when $\pi_0$ is close to 1).  
  \item The two test statistics (the Bayes factor and min-$p$) display very similar powers in gene-level testing, however the Bayes factor performs slightly better.  
\end{enumerate}

\begin{table}[ht]
\begin{center}
\begin{tabular} { c  c   c c  c  c c  c  c c  c   c c  c  c  c  c }
\hline

   ~ & ~& \multicolumn{15}{c}{FDP} \\
   \cline{3-17} 
  $\pi_0$ & ~& \multicolumn{2}{c}{B-H ($p_{minp}$) } & ~ & \multicolumn{2}{c}{B-H ($p_{bf}$) } & ~ & \multicolumn{2}{c}{ Storey ($p_{minp}$) } & ~ & \multicolumn{2}{c}{Storey ($p_{bf}$) } & ~ & ~ & ~ & ~\\
  \cline{3-4} \cline{6-7} \cline{9-10} \cline{12-13} 
~ & ~ & 500 & 5000 & ~ & 500 & 5000 & ~  & 500 & 5000 & ~ & 500 & 5000 & ~ & QBF & ~ & EBF \\
\hline
\hline
0.95 & &  $0.000^\dagger$ & 0.046 &  & $0.000^\dagger$ & 0.043 & & $0.000^\dagger$ & 0.046 &  & $0.000^\dagger$ & 0.043 &   &0.050 &   & 0.035 \\
0.85 & ~ &  0.033 & 0.043 & ~ &  0.031 & 0.043 & ~ & 0.033 & 0.047 & ~ & 0.031 & 0.048 & ~ &    0.039 & ~&  0.028 \\
0.75 & ~ &  0.031 & 0.036 & ~ &   0.029 & 0.036 &~&  0.042 & 0.042 & ~ & 0.041 & 0.043 & ~ &       0.031 & ~ & 0.021 \\
0.65 & ~ & 0.028  & 0.032 & ~ &   0.026 & 0.032 &~&  0.038 & 0.041 & ~ & 0.040 & 0.042 & ~ &       0.029 & ~ & 0.020 \\
0.55 & ~ &  0.025 &  0.028 &~&    0.025 & 0.028 &~&  0.038 & 0.039 &~&  0.038 & 0.038  &~&      0.025 &~& 0.017 \\
0.45 &~&  0.021 & 0.022  &~&  0.021 & 0.022 &~&  0.032 & 0.034 &~& 0.033  & 0.034  &~&  0.022  &~& 0.014 \\
0.35 &~&  0.018 & 0.018 &~&   0.017 & 0.018 &~&  0.029 & 0.030 &~& 0.029  & 0.030  &~& 0.020 &~& 0.013 \\
0.25 &~& 0.012 & 0.013 &~&   0.011 & 0.012  &~& 0.023 & 0.024 &~&  0.024 & 0.024   &~&  0.016 &~& 0.009 \\
0.15 &~& 0.007 & 0.008 &~&   0.007 & 0.008 &~& 0.016  & 0.016 &~&  0.016 & 0.017  &~& 0.011 &~& 0.006 \\
\hline
\end{tabular}
\caption{\label{fdp.sim2.tbl}Realized FDPs yielded by the B-H,  Storey, QBF, and EBF procedures in simulation study II. Each entry is obtained by averaging over the 20 simulated data sets. The entries annotated by ``$\dagger$" indicate that the corresponding method fails to reject any hypothesis. }

\end{center}
\end{table}

\begin{table}[ht]
\begin{center}
\begin{tabular} { c  c   c c  c  c c  c  c c  c   c c  c  c  c  c }
\hline

   ~ & ~& \multicolumn{15}{c}{FNP} \\
   \cline{3-17} 
  $\pi_0$ & ~& \multicolumn{2}{c}{B-H ($p_{minp}$) } & ~ & \multicolumn{2}{c}{B-H ($p_{bf}$) } & ~ & \multicolumn{2}{c}{ Storey ($p_{minp}$) } & ~ & \multicolumn{2}{c}{Storey ($p_{bf}$) } & ~ & ~ & ~ & ~\\
  \cline{3-4} \cline{6-7} \cline{9-10} \cline{12-13}   
~ & ~ & 500 & 5000 & ~ & 500 & 5000 & ~  & 500 & 5000 & ~ & 500 & 5000 & ~ & QBF & ~ & EBF \\
\hline
\hline
0.95 &~&  $0.050^\dagger$ & 0.034 &~&  $0.050^\dagger$ & 0.034 &~&  $0.050^\dagger$ & 0.034 &~& $0.050^\dagger$ & 0.034 & ~& 0.034 &~& 0.034 \\
0.85 &~&  0.104 & 0.101 &~&  0.103 & 0.100 &~& 0.104 & 0.100 &~& 0.103 & 0.099 &~& 0.100 &~& 0.103 \\
0.75 &~&  0.173 & 0.169 &~& 0.171 & 0.167 &~&  0.169 & 0.168 &~& 0.167 & 0.165 &~& 0.168 &~& 0.172 \\
0.65 &~& 0.247 & 0.243 &~& 0.244 & 0.240 &~&  0.242 & 0.240 &~&  0.238 & 0.236 &~& 0.241 &~& 0.247 \\
0.55 &~& 0.329 & 0.325 &~& 0.325 & 0.321 &~&  0.321 & 0.319 &~& 0.316 & 0.314 &~& 0.322 &~&  0.330 \\
0.45 &~& 0.417 & 0.414 &~& 0.411 & 0.409 &~&    0.408 & 0.406 &~& 0.401 & 0.399 &~&  0.409 &~&  0.418 \\
0.35 &~&  0.516 & 0.515 &~& 0.511 & 0.509 &~&    0.505 & 0.503 &~& 0.497 & 0.495 &~&  0.507 &~& 0.517 \\
0.25 &~&  0.631 & 0.629 &~& 0.626 & 0.623  &~&   0.616 & 0.615 &~& 0.608 & 0.607 &~& 0.618 &~& 0.630 \\
0.15 &~&  0.761 & 0.760 &~& 0.757 & 0.755  &~&   0.746 & 0.745 &~& 0.739 & 0.738 &~& 0.748 &~& 0.759 \\
\hline
\end{tabular}
\caption{\label{fnp.sim2.tbl}Realized FNPs obtained by the B-H,  Storey, QBF, and EBF procedures in simulation study II. Each entry is obtained by averaging over  the 20 simulated data sets. The entries annotated by ``$\dagger$" indicate that the corresponding method fails to reject any hypothesis and the FNP achieves the maximum value.} 
\end{center}
\end{table}

Although all the procedures yield extremely similar statistical performance, they have striking differences in computational efficiency. We benchmark the running time of a complete analysis of a single simulated data set using various procedures. All of the procedures tested are implemented in the C++ programming language and run on a machine with an 8-core Intel Xeon 2.13GHz CPU. We present the results in Table \ref{comp.time.sim2.tbl}. Given required $p$-values or Bayes factors, the computations for various FDR control procedures are mostly trivial, the main difference in computational time is primarily attributed to the cost in permutations.

\begin{table}[ht]
\begin{center}
\begin{tabular} { c | c | c | c }
\hline
EBF & QBF (100) & B-H/Storey (500)  & B-H/Storey (5000) \\ 
\hline
\hline
 0m43s & 3m34s & 11m56s & 114m34s \\
 \hline
\end{tabular}\caption{\label{comp.time.sim2.tbl} Computational time measured for various FDR control procedures in simulation study II. Numbers in parenthesis indicate the number of permutations performed for each procedure.} \end{center}
\end{table}

\section{Real Data Application: Multi-tissue eQTL Discovery} 

In this section, we demonstrate the proposed statistical approach in a real data application of mapping eQTL across multiple tissues. 
The aim of the multi-tissue eQTL mapping is to investigate the commonality and the specificity of the eQTLs in different tissue environments.
Comparing to the single-tissue eQTL study described in the simulation study, expression levels for each gene are  collected in multiple tissue/cell types from the same set of unrelated individuals, and a joint eQTL analysis across all tissues is required.
Within each tissue, we assume the same statistical model as the linear model (\ref{sim2.model}), however across tissues, the residual errors are generally assumed to be correlated \citep{Wen2014a}.

Our primary goal is still to identify genes harboring causal eQTLs (i.e., eGenes): let $\beta_{i_{jt}}$ denote the association coefficient between expressions of gene $i$ and its $j$-th genetic variant in tissue $t$. For each gene $i$, we are interested in testing the following hypothesis:
\begin{equation} \label{multi.tissue.test}
  H_0: \beta_{i_{jt}} =0, \forall \, j, t ~~~vs. ~~~ H_1: \mbox{ some } \beta_{i_{jt}} \ne 0. 
\end{equation}  
Clearly, the problem is even more challenging when considering the potential association patterns across multiple tissues under the {\em alternative} scenario.
First, multi-tissue data allow identifying association signals that might be modest but consistent across tissues. To achieve this identification, it requires a framework analogous to a meta-analysis that jointly models potential associations across tissue types. Secondly, it is known that a proportion of eQTLs are only active in specific tissue types (i.e., $\beta_{i_{jt}}  \ne 0$ only for some $t$). To account for this, we must explicitly address the configurations of possible association activities across multiple tissues. 

In  \cite{Dimas2009}, the authors used a naive gene-level test statistic $T_i = \min_{j,t} p_{i_{jt}}$, where each $p_{i_{jt}}$ is the $p$-value from a single association test for each gene-genetic variant pair in each tissue type. The $p$-value of the overall test statistic, $T_i$, must be evaluated in a very complicated permutation scheme to account for intra-individual correlations of gene expression levels and the LD in genetic variants. Apparently, this naive test statistic does not fully capture the alternative scenarios discussed above and is expected to lose power. 
In brief, to the best of our knowledge, {\em there is no frequentist approach that can produce appropriate $p$-values to test (\ref{multi.tissue.test})}.

Recently, \cite{Flutre2013} proposed a full parametric Bayesian model to explicitly describe all possible alternative scenarios and derived a Bayes factor to efficiently test eGenes,  taking advantage of Bayesian model averaging. 
More specifically, for eQTL data collect from $T$ different tissues, they consider all $2^T-1$ possible configurations of association activities for each potential associated SNP (e.g., a configuration $(1,1,...,1)$ indicates the eQTL is active in all tissues). For each given configuration, a Bayes factor can be computed analytically accounting for the similarity and heterogeneity of the genetic effect sizes ($\beta_{i_{jt}}$'s) of each active association in different tissues \citep{Wen2014b}. 
The probability weights on each type of the total $(2^T -1)$ non-zero configuration model are then estimated using an EM algorithm by pooling information across all available gene-SNP data across the genome. 
Given the resulting weights on configurations and the ``(at most) one causal association per {\it cis} region" assumption, a weighted Bayes factor is evaluated using Bayesian model averaging. 
Note that the weighted Bayes factor itself is a well-defined Bayes factor and can be plugged into the proposed EBF and QBF procedures.  
\cite{Flutre2013} convincingly demonstrated that the resulting Bayes factor is much more powerful than the naive test statistic $T_i=\min_{j,t} p_{i_{jt}}$.
Nevertheless, to apply a multiple testing procedure for identifying eGenes, they had to rely on permutation procedures to obtain $p$-values by treating the resulting Bayes factors as the frequentist test statistics.

We apply both the EBF and the QBF procedures to a real multiple tissue eQTL data set  that was originally published in \cite{Dimas2009} and re-analyzed by \cite{Flutre2013}. In the experiment, 75 unrelated Western European individuals are genotyped. Expression levels from this set of individuals were measured genome-wide in primary fibroblasts, Epstein-Barr virus-immortalized B cells (LCLs), and T cells. The expression data went through quality control and normalization steps (by \cite{Dimas2009}), and a subset of 5,012 genes that were deemed highly expressed in all three cell types is selected for eQTL analysis. 

Based on the Bayes factor derived in \cite{Flutre2013}, the QBF (in which $\hat \pi_0$ is estimated with only 100 permutations) and EBF  procedures identify 1002  and 927 eGenes, respectively, at a Bayesian FDR level of 5\%. In comparison, \cite{Flutre2013} identified 1012 eGenes by applying the Storey procedure using the permutation $p$-values of the Bayes factors (with 10,000 permutations) also at an FDR level of 5\%. 
Although the three methods yield similar results, the EBF and the QBF procedures achieve dramatic computational time reduction by avoiding large amounts of permutations. Finally, \cite{Flutre2013} noted that Storey's procedure based on permutation $p$-values of the naive $T_i$ statistics (with 10,000 permutations) only identified 627 eGenes. Among those eGenes, 592 genes(or $94\%$) are also identified by the EBF and QBF approaches.

\section{Discussion}

We have introduced a Bayesian FDR control procedure with Bayes factors that is robust to misspecifications of alternative models. This feature should provide peace of mind for practitioners who are attempting parametric Bayesian models in multiple hypothesis testing.   
Nevertheless, within our framework, the model specification still dictates the overall performance, e.g., a badly designed alternative model would have very little power and would therefore be useless. 
Our central message throughout this paper has been that various FDR control procedures have little practical difference if the same or similar test statistics are applied; however, our proposed procedure encourages well-designed parametric modeling approaches to obtain more powerful test statistics. 

Another advantage of our approach is its computational efficiency, especially when compared with the alternative method's reliance on permutation $p$-values. However, this statement assumes the Bayes factors can be easily evaluated. 
Many previous and ongoing works have been devoted to efficient evaluation and approximation of Bayes factors in various model systems. In particular, \cite{Johnson2005} demonstrated that almost all the well-known frequentist test statistics (e.g., $z, F$, and $\chi^2$ statistics) can be ``translated" into corresponding Bayes factors analytically, and \cite{Wen2014a} demonstrated that Bayes factors can be analytically approximated in a general linear model system that includes multiple linear regression, multivariate linear regression models, and many others as special cases.

Our proposed procedure bears certain similarities to the local fdr procedure \citep{Efron2001,Sun2007}. However, it should be noted that the two procedures differ significantly in the following two aspects. First, our procedure utilizes Bayes factors as test statistics whereas local fdr procedure relies on $z$-statistics. We note that in complex parametric modeling settings, $z$-statistic is not always straightforwardly defined (e.g., in our eQTL mapping examples). 
Second, the local fdr procedure is an empirical Bayes approach: it plugs in the MLE of $\pi_0$ instead of performing full Bayesian inference on $\pi_0$. As discussed in this paper, this approach can be susceptible to model misspecifications. In contrast, our proposed approach is robust to model specification, and philosophically is fully Bayesian by viewing $\hat \pi_0$ as a conservative approximation (by a point mass) to the posterior distribution of $\pi_0$.

\section{Acknowledgments}

We thank Debashis Ghosh, Matthew Stephens and Timothee Flutre for discussion and feedbacks. We are grateful for the insightful comments from the two anonymous reviewers. 
This work is supported by the NIH grant R01 MH101825 (PI: M.Stephens).

\newpage
\appendix{Appendix}

\section{Proof of Proposition 1} \label{thm1.proof}
\begin{proof} 
Suppose that the true generative distributions of $\Yv_i$ under the null and alternative model are given by $p^{0^*}_i$ and $p^{1^*}_i$, respectively. As sample size $n_i$ is sufficiently large for each test, the true Bayes factor, $\BF_i^*$, has the following properties
\begin{equation}
  \label{con1}
  \begin{aligned}
     &\BF_i^* \xrightarrow{P} 0,~\mbox{if } \Yv_i \sim  p^{0^*}_i, \\
     &~~~~~~~~~\mbox{and} \\
     & \BF_i^* \xrightarrow{P} \infty,~\mbox{if } \Yv_i \sim  p^{1^*}_i.\\  
  \end{aligned}
\end{equation}
Assumption 1, in contrast, only requires that the assumed Bayes factors satisfy
\begin{equation}
   \label{con2}
   \BF_i \xrightarrow{P} 0,~\mbox{if } \Yv_i \sim  p^{0^*}_i.
\end{equation} 
Under the conditions stated in Assumption 2,  
$\Pr \left( \hat \pi_0 \ge \pi_0 \mid \pi_0 \right) \to 1$
implies that $\Pr \left( \hat \pi_0 \ge \pi_0 \mid \Yva \right) \to 1$ for an arbitrary prior distribution on $\pi_0$. Let $\epsilon := \Pr \left( \hat \pi_0 < \pi_0 \mid \Yva \right)$. Then, it follows that 
\begin{equation}
   \label{con3}
   \epsilon  \xrightarrow{P}  0,~\mbox{as the number of null tests is sufficiently large.}
\end{equation}
Importantly, the large number of the null tests is required for ensuring the upper bound property of $\hat \pi_0$.

Consequently,
\begin{equation}
  \label{con4}
  \begin{aligned}
  \Pr(Z_i = 0 \mid \Yva) &= \int \frac{\pi_0}{\pi_0 + (1-\pi_0) \BF_i^*} \, p(\pi_0 \mid \Yv_i)\,d\,\pi_0 \\  
                       &= \int_{\pi_0 \le \hat \pi_0} \frac{\pi_0}{\pi_0 + (1-\pi_0) \BF_i^*} \, p(\pi_0 \mid \Yv_i)\,d\,\pi_0 + \int_{\pi_0 > \hat \pi_0} \frac{\pi_0}{\pi_0 + (1-\pi_0) \BF_i^*} \, p(\pi_0 \mid \Yv_i)\,d\,\pi_0  \\
                       &\le \frac{\hat \pi_0}{\hat \pi_0 + (1- \hat \pi_0) \BF_i^*} + \epsilon                   
   \end{aligned}
\end{equation}
By (\ref{con1}), as $n_i \to \infty$, 
\begin{equation}
    \frac{\hat \pi_0}{\hat \pi_0 + (1- \hat \pi_0) \BF_i^*}  \xrightarrow{P} I\{\Yv_i \sim p^{0*}_i \} + 0 \cdot I\{\Yv_i \sim p^{1*}_i \} 
\end{equation}
whereas by (\ref{con2}),
\begin{equation}
    \begin{aligned}
    &\frac{\hat \pi_0}{\hat \pi_0 + (1- \hat \pi_0) \BF_i}  \xrightarrow{P} 1,  ~\mbox{ if } ~ \Yv_i \sim p^{0*}_i,  \\
     &\frac{\hat \pi_0}{\hat \pi_0 + (1- \hat \pi_0) \BF_i} \ge 0, ~\mbox{ if } ~ \Yv_i \sim p^{1*}_i.
   \end{aligned}  
\end{equation}
Hence, 
\begin{equation}
  \lim_{n \to \infty} \Pr \left( \frac{\hat \pi_0}{\hat \pi_0 + (1- \hat \pi_0) \BF_i} \ge \frac{\hat \pi_0}{\hat \pi_0 + (1- \hat \pi_0) \BF_i^*} \right) = 1, 
\end{equation}
and by (\ref{con3}) and (\ref{con4}), each individual test satisfies
\begin{equation}
  \label{thm1.rst1}
  \lim_{n_i \to \infty}   \Pr \left( (1 - \hat v_i)  \ge \Pr(Z_i = 0 \mid \Yv_i) \right) = 1 
\end{equation}

The decision rule stated in the Proposition 1 yields a true Bayesian FDR
\begin{equation}
  \overline{\FDR} = \frac{\sum_{i=1}^m \delta^\dagger_i \Pr (Z_i =0 \mid \Yv_i)}{D^\dagger \vee 1}.
\end{equation}    
By (\ref{thm1.rst1}), it is clear that
\begin{equation}
  \lim_{n \to \infty} \Pr \left( \BFDR \le \frac{\sum_{i=1}^m \delta^\dagger_i (1-\hat v_i)}{D^\dagger \vee 1} \right) = 1.
\end{equation}
Therefore, the Bayesian FDR can be consistently controlled. Furthermore, controlling $\frac{\sum_{i=1}^m \delta^\dagger_i (1-\hat v_i)}{D^\dagger \vee 1} \le \alpha$ ensures that
\begin{equation}
   \FDR = {\rm E}(\overline{\FDR}) \le {\rm E}\left(\frac{\sum_{i=1}^m \delta^\dagger_i (1-\hat v_i)}{D^\dagger \vee 1} \right) \le \alpha,
\end{equation}
because $\BFDR$ is obviously bounded.
\end{proof}

\section{Proof of Lemma 2}\label{lem2.proof} 
\begin{proof}

Let $(\BF_{(1)},\dots,\BF_{(m)})$ denote the  order statistics from the $m$ Bayes factors that are all generated from the respective null hypotheses. 
Let $M_j := \frac{1}{j} \sum_{i=1}^j \BF_{(i)}$ denote the partial sample mean computed by the EBF procedure. 
Note that the sequence $M_1, M_2, \dots $ is monotonically non-decreasing. Furthermore, by the law of large numbers and the result of Lemma 1, it follows that
\begin{equation}
 M_m  \xrightarrow{P} 1,
\end{equation}
for sufficiently large $m$.

In the case that $d_0 < m$, it must be true that 
$$  1 \le M_{d_0+1} \le M_m \xrightarrow{P} 1. $$
Because the truncated sample mean from the order statistics $(\BF_{(d_0+1)}, ... , \BF_{(m)})$ converges to a quantity that is strictly greater than 1, their contribution to the overall sample mean, $M_m$, must be negligible, i.e.,
$$ \frac{m - d_0}{m} \, \left( \frac{1}{m-d_0}\sum_{j=d_0+1}^m \BF_{(j)} \right) \xrightarrow{P} 0. $$ 
Taken together, we have shown that  
\begin{equation}
   \hat \pi_0 \, = \, \frac{d_0}{m} \, \xrightarrow{P} \, 1.
\end{equation}

\end{proof}

\section{Proof of Proposition 2}\label{prop1.proof}
\begin{proof}
In the general mixture case, let $\mathcal{S}^0$ denote the subset of Bayes factors whose data are generated from the null models. Based on Lemma 2, applying the EBF procedure on $\mathcal{S}^0$ results in an estimate 
\begin{equation} \label{c.1}
\frac{d_{\mathcal{S}^0}}{ |\mathcal{S}^0|} \xrightarrow{P} 1 ,
\end{equation}
where $|\mathcal{S}^0|$ denotes the cardinality of $\mathcal{S}^0$.
In the mixed samples, (\ref{c.1}) suggests that 
\begin{equation}
  M_{d_{\mathcal{S}^0}} = \frac{1}{d_{\mathcal{S}^0}} \, \sum_{j=1}^{d_{\mathcal{S}^0}} \BF_{(j)} \le 1.
\end{equation}  
The LHS should be strictly less than 1 if there exists small values of Bayes factors from the alternative models.

Because the EBF procedure finds the largest subset whose sample mean is less than 1, it must hold true that 
\begin{equation}
  \Pr( d_0 \ge d_{\mathcal{S}^0} \mid \pi_0 ) \to 1,
\end{equation}
and thus we conclude that 
\begin{equation}
  \Pr(\hat \pi_0 \ge \pi_0 \mid \pi_0 ) \to 1.
\end{equation}

\end{proof}
  
\section{Proof of Corollary 1}\label{corollary1.proof}
\begin{proof}
Given a pre-defined FDR level $\alpha$, the rejection threshold on the estimated false discovery probabilities is given by
$$ t^\dagger_{\alpha} =   \arg\min_t  \left( \frac{\sum_{i=1}^m   \delta_i^\dagger(t) (1-\hat v_i )}{D^\dagger(t) \vee 1} \le \alpha \right).$$
Equivalently when there is at least one rejection,  the above rejection threshold implies that the rejection set $\Omega := \{i: \hat v_i > t^\dagger_{\alpha}\}$ is the largest set such that 
$$ \frac{\sum_{ i \in \Omega} \Pr(Z_i =0 \mid \Yv_i, \hat \pi_0)}{|| \Omega || } \le  \alpha, $$
where $\Pr(Z_i = 0 \mid \Yv_i, \hat \pi_0) = (1-\hat v_i)$ and $|| \Omega || = D^\dagger(t_{\alpha}^\dagger)$ denotes the cardinality of the set $\Omega$. That is, 
the average estimated false rejection probability in the rejection set should be $\le \alpha$.  
Consequently, it implies that if the $i$-th test yields $\Pr(Z_i = 0 \mid \Yv_i, \hat \pi_0) \le \alpha$, it must be included in the rejection set (because $\Omega$ is the largest such set). 
Therefore, to prove the corollary,  it is sufficient to show that $\Pr(Z_i=0 \mid \Yv_i, \hat \pi_0) \le  \alpha$, provided that  $\BF_i \ge \frac{m}{\alpha}$.

Applying the EBF procedure, a single Bayes factor with the value exceeding $m$ leads to 
\begin{equation}
  \hat \pi_0 \le 1- \frac{1}{m}.
\end{equation}
This implies
\begin{equation}
  \Pr(Z_i=0 \mid \Yv_i, \hat \pi_0) \le \frac{ 1- \frac{1}{m} }{(1- \frac{1}{m}) + \frac{1}{m} \BF_i} \le  \frac{m-1}{m+(m-1) \alpha} \cdot \alpha < \frac{m-1}{ m -1 + m \alpha} \cdot \alpha < \alpha
\end{equation}
\end{proof}

\newpage

\bibliographystyle{natbib}  
\bibliography{BFDR}

\end{document}